\documentclass[a4paper,fleqn,usenatbib,useAMS,usefloat,onecolumn ]{mnras}
\usepackage{mathptmx}
\usepackage[T1]{fontenc}
\usepackage{ae,aecompl}

\usepackage{graphicx}	
\usepackage{amsmath}	
\usepackage{amssymb}	
\usepackage{subfig}
\usepackage{ulem}
\usepackage{epstopdf}
\usepackage{booktabs}
\usepackage{array}
\usepackage{caption}
\title[SB 796: a high-velocity RRc star]
{SB 796: a high-velocity RRc star}
\author[Gomel et al.]{
\newauthor{Roy Gomel,$^1$ Sahar Shahaf,$^1$ Tsevi Mazeh,$^1$ Simchon Faigler,$^1$} 
\newauthor{Lisa A. Crause,$^{2}$ Ramotholo Sefako,$^{2}$ Damien Segransan,$^{3}$} 
\newauthor{Pierre F.L. Maxted$^{4}$ and Igor Soszy{\'n}ski$^{5}$}
\\
$^{1}${ School of Physics and Astronomy, Raymond and Beverly Sackler Faculty of Exact Sciences,}\\
{\ \ Tel Aviv University, Tel Aviv  69978, Israel}\\
$^{2}${South African Astronomical Observatory, PO Box 9, Observatory 7935, South Africa}\\
$^{3}${Observatoire de Geneve, 51 ch des Maillettes, CH-1290 Sauverny, Switzerland}\\
$^{4}${Astrophysics Group, Keele University, Staffordshire, ST5 5BG, UK}\\
$^{5}${Warsaw University Observatory, Al Ujazdowskie 4, 00-478 Warszawa, Poland}\\
}
\date{Accepted XXX. Received YYY; in original form ZZZ}

\pubyear{2018}

\begin{document}
\label{firstpage}
\pagerange{\pageref{firstpage}--\pageref{lastpage}}
\maketitle

\begin{abstract}

We report here on a detailed study of a c-type RR Lyrae variable (RRc variable), SB 796, serendipitously discovered in a search of the WASP public data for stars that display large photometric periodic modulation. 
SB 796 displays a period of $P = 0.26585$ d and semi-amplitude of $\sim$ 0.1 mag. 
Comparison of the modulation shape and period with the detailed analysis of LMC variables indicates that SB 796 is an RRc variable. 
{\it Gaia} DR2 classification corroborated our result. 
 Radial-velocity (RV) follow-up observations revealed a periodic variation
consistent with a sine modulation, with a semi-amplitude of $5.6\pm0.2$ km/s, and a minimum  at phase of maximum brightness. Similar amplitude and phase were previously seen in other RRc variables.
The stellar averaged RV is $\sim 250$ km/s, turning SB 796 to be a high-velocity star, while its present position, as derived from the {\it Gaia} astrometry, is 
at $\sim 3.5$ kpc below the Galactic plane.
Integration of the stellar Galactic motion shows that 
SB 796 oscillates at a range of 0.5--20 kpc Galacto-centric distance, passing near the Galactic center about three times in 1 Gyr. The Galactic radial motion takes SB 796 up and down the plane to a scale height of $\sim 10$ kpc. During its $\sim10$ Gyrs estimated life time, SB 796 therefore passed $\sim 30$ times near the Galactic center. 

\end{abstract}

\begin{keywords}
{
stars: variables: RR Lyrae -- stars: Population II -- Galaxy: kinematics and dynamics -- Galaxy: halo
}
\end{keywords}
\section{Introduction}

RR Lyrae stars are relatively low-mass ($0.6$--$0.8M_{\odot}$)
pulsating stars that undergo core helium burning on the
horizontal branch \citep{smith09}.
Being bright Population II objects that pulsate in regular modes, they can be used as tracers of Galactic structure and history \citep[e.g.,][]{akhter12, pietrukwicz15, cohen17, ablimit18}. 
As pointed out by \cite{soszy03,soszy09}, RR Lyrae stars can be identified by their pulsation periods, light-curve shapes, luminosities and colors. 
Therefore, the accumulating photometric databases of billions of stars, the {\it Gaia} database in particular 
\citep[e.g.,][]{Riello18, evans18},
 open up an opportunity to identify a large number of new RR Lyrae stars that will enable us to study in details their dynamical history during the early history of the Galaxy \citep{layden95}. 

We report here on a detailed study of one c-type RR Lyrae variable (RRc variable) that was serendipitously discovered in a search for stars with massive companions that display large photometric ellipsoidal variations (Mazeh et al., in preparation). Searching the WASP public data\footnote{https://wasp.cerit-sc.cz/form} \citep[][]{pollacco06, butters10} for periodic photometric variables we came across a stellar candidate that displayed a brightness modulation with a period of  $\sim 0.25$ d. The star appears in the compilation of 
A
stars located near the south Galactic pole \citep{slettebak71} as SB 796 (see  Table~\ref{table:Sys_Parameters}), putting its present position well below the Galactic plane. 

Radial-velocity (RV) follow-up observations,
 performed with the SpUpNIC spectrograph\footnote{http://www.saao.ac.za/science/facilities/instruments/spupnic-spectrograph-upgrade-newly-improved-cassegrain} \citep{crause16} at the 
South African Astronomical Observatory (SAAO), revealed a small amplitude RV modulation, indicating that the photometric modulation is not caused by the ellipsoidal effect. Further RV measurements with higher accuracy, obtained with the CORALIE spectrograph \citep{queloz00,pepe02} at the Swiss Euler telescope, were consistent with the photometric period with a small amplitude, confirmed the identification of SB 796 as a stellar pulsator.
Comparison with the detailed analysis of LMC variables by \citet{soszy09,soszy16}
attested to the classification of SB 796 as an RRc variable.
When we prepared our results for publication, {\it Gaia} \citep{prusti16} released its DR2 results \citep{brown18}, 
 including classification of 40,380 RRc variables, together with 100,404 RR Lyrae stars of other types \citep{clementini18}, 
and indeed SB 796 was listed as an RRc variable.
 Consequently, the {\it Gaia} time series observations of SB 796 became available \citep{holl18} so we could include them in the analysis of the photometry.

The obtained RVs displayed an averaged radial-velocity of $\sim 250$ km/s, indicating that SB 796 is a high-velocity star \citep[e.g.][]{schuster89}, moving away from the Galactic plane. {\it Gaia}'s five-parameter astrometry \citep{lindegren18} of SB 796, combined with our RVs, enabled us to study the Galactic motion of SB 796 in the past and into the future.
The star is moving in a Galactic radial motion, oscillating at a range of 
$\sim0.5$--$20$ kpc Galacto-centeric distance in $\sim0.3$ Gyr.  


Section~\ref{sec:photometry} describes our analysis of the WASP and  {\it Gaia} photometry of SB 796 and Section~\ref{sec:RV} presents our RV observations. 
Section~\ref{sec:classification} shows that the characteristics of SB 796 are typical of an RRc variable and Section~\ref{sec:galactic} analyses its Galactic motion. Finally, Section~\ref{sec:summary} summaries our results.

\begin{table}
	\centering
	\begin{tabular}{l c} 
		\hline
		Coordinates (J2000) & 02$^h$ 02$^m$ 29$^s$.05 -40$^o$ 22' 25''.9$^a$ \\
		Spectral Type 		& A0$^b$ \\
		B 					& 13.1$^b$ \\
		V 					& 13.4$^b$ \\
		\hline \\
		$^a$WASP public data \\
		$^b$\citet{slettebak71} \\
	\end{tabular}
	\caption{Basic information of SB 796.}
	\label{table:Sys_Parameters}
\end{table}

\section{Photometric modulation}
\label{sec:photometry}

The WASP project \citep{pollacco06} observed SB 796 during 2006--2014, 
obtaining 49,804 measurements spreading over six observing seasons, with a mean range of $\sim$ 150 d per season. The observations of the first four seasons (2006--2012) were taken with a broadband, 400--700nm filter (Visual-WASP or $V_W$ band) and 200-mm, f/1.8 lenses . The data of the last two seasons (2013--2014) was taken with an r' filter, using 85-mm, f/1.2 lenses \citep{smith14}. 

\begin{figure*}
	\centering
	\includegraphics[width=16cm, height=15cm]{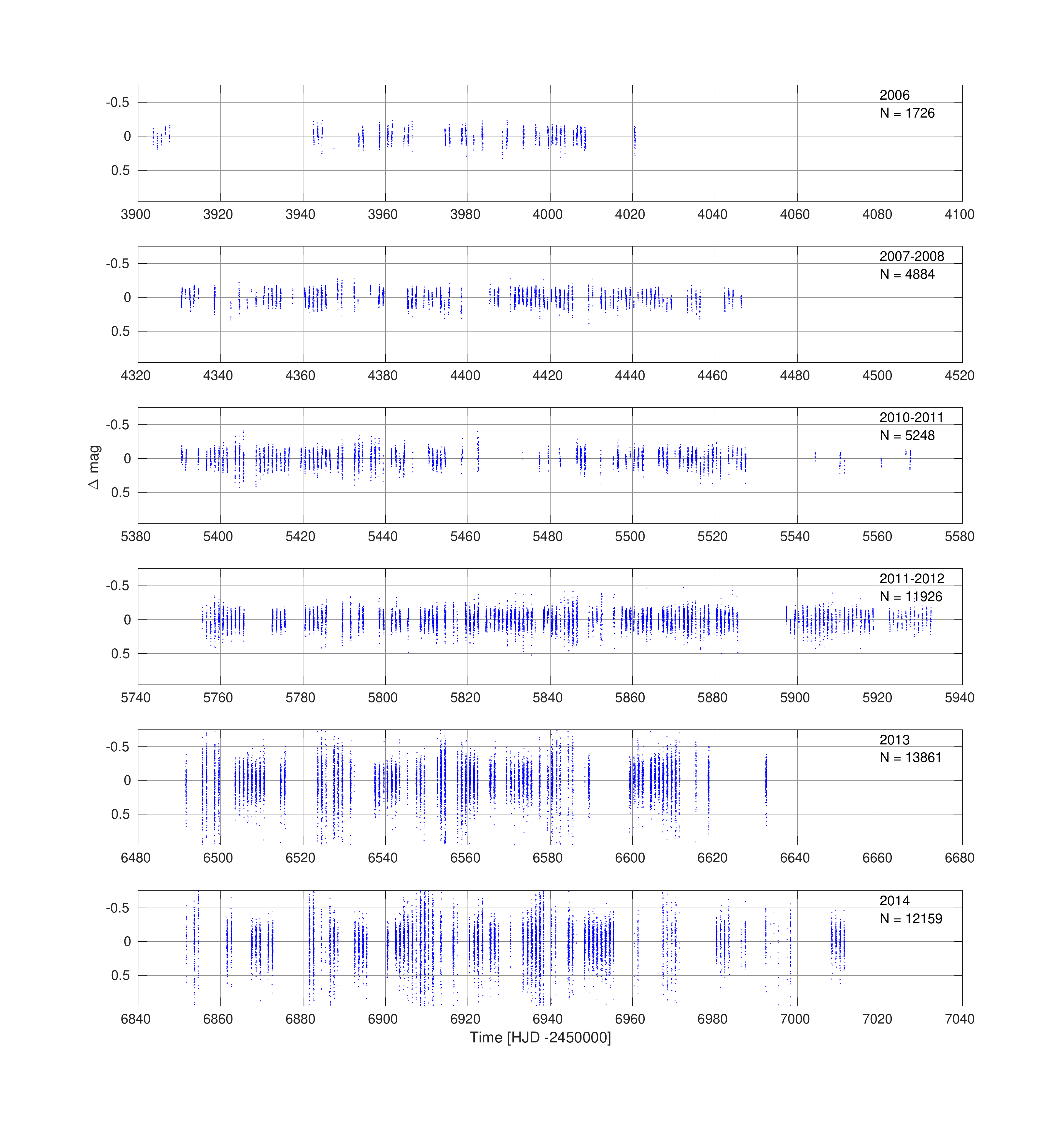}
	\caption{
WASP photometry of SB 796, spans over $\sim$3000 d, divided into six seasons. Seasons 1--4 (2006--2012) were observed in the $V_W$ band while seasons 5--6 (2013--2014) in the r' band. For each season, the data were median subtracted.
	} 
	\label{fig:WASP_lc}
\end{figure*}

Typically 50 measurements of SB 796 were taken each night, during about 5 hours, with a median cadence of $\sim$ 40 s. The WASP light curve is plotted in Fig.~\ref{fig:WASP_lc}. Seasons 5--6 display a larger photometric scatter because their data was taken with smaller aperture (85-mm) lenses.

\begin{figure*}
	\centering
	\includegraphics[width=12cm, height=6cm]{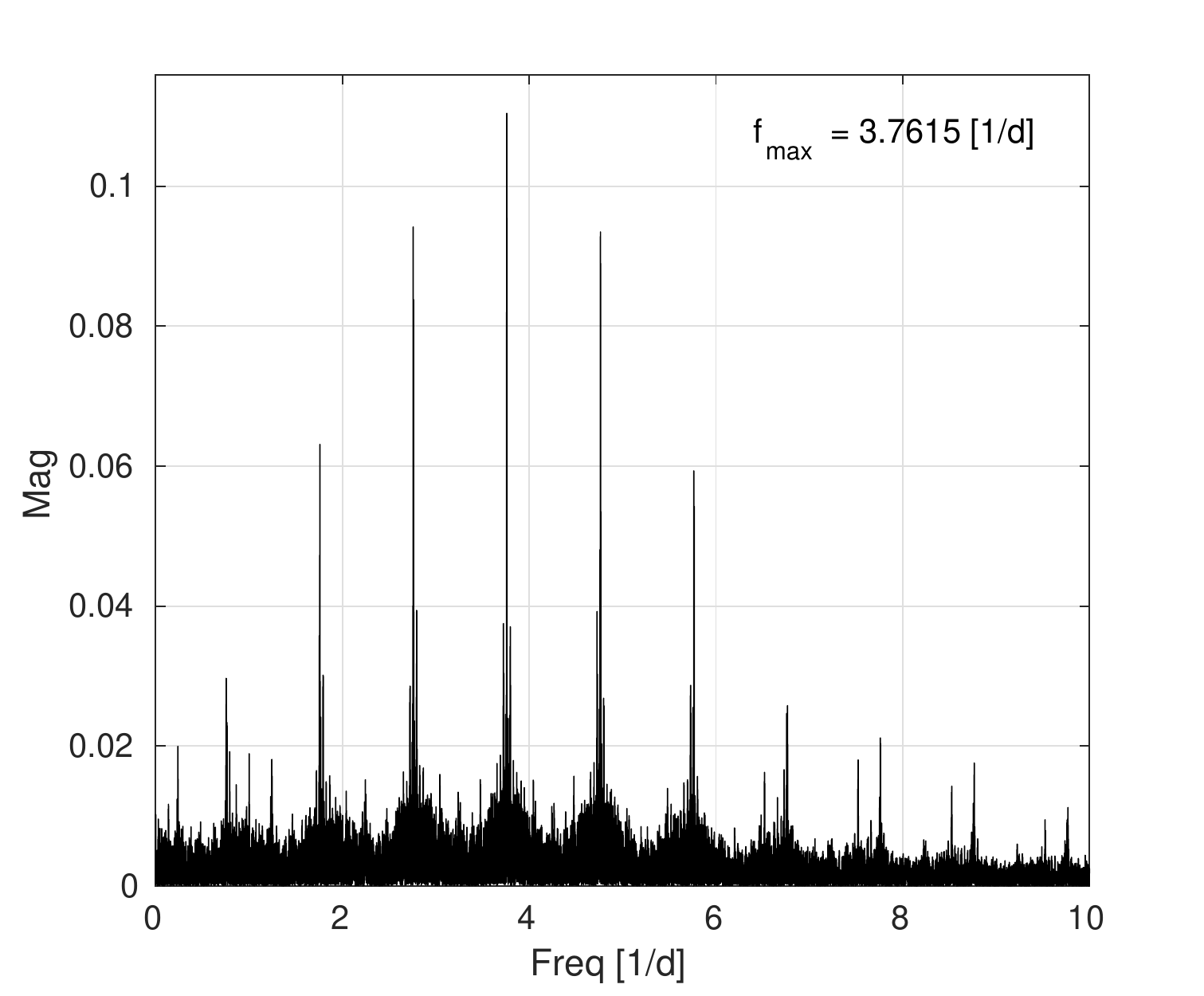}
	\caption{
		Magnitude spectrum of WASP $V_W$ photometry as a function of frequency, with a peak at $\sim$ 0.266 d. The peak height of $\sim$ 0.11 mag indicates the semi-amplitude of the first harmonic.}
	\label{fig:WASP_PS}
\end{figure*}

The magnitude spectrum of the $V_W$ light curve is shown in Fig.~\ref{fig:WASP_PS}, with a central peak corresponding to a period of P $\sim$ 0.266 d. The other peaks, which appear symmetrically on both sides of the main peak, are due to the strong daily aliases induced by the window function of  the data. 

We performed a fine search around the period of the main peak by fitting a 3-harmonic model to the observed magnitudes in the form of \begin{equation}
A_0 + \sum_{n=1}^3 A_n \cos\big(\frac{2 \pi n}{P} t+\phi_n\big),
\label{eq:harmonics}
\end{equation} for each season of WASP data set. The magnitude uncertainties were normalized to get a reduced $\chi^2_{red}$ = 1 per season. 

\begin{figure*}
	\centering
	\includegraphics[width=12cm, height=7.5cm]{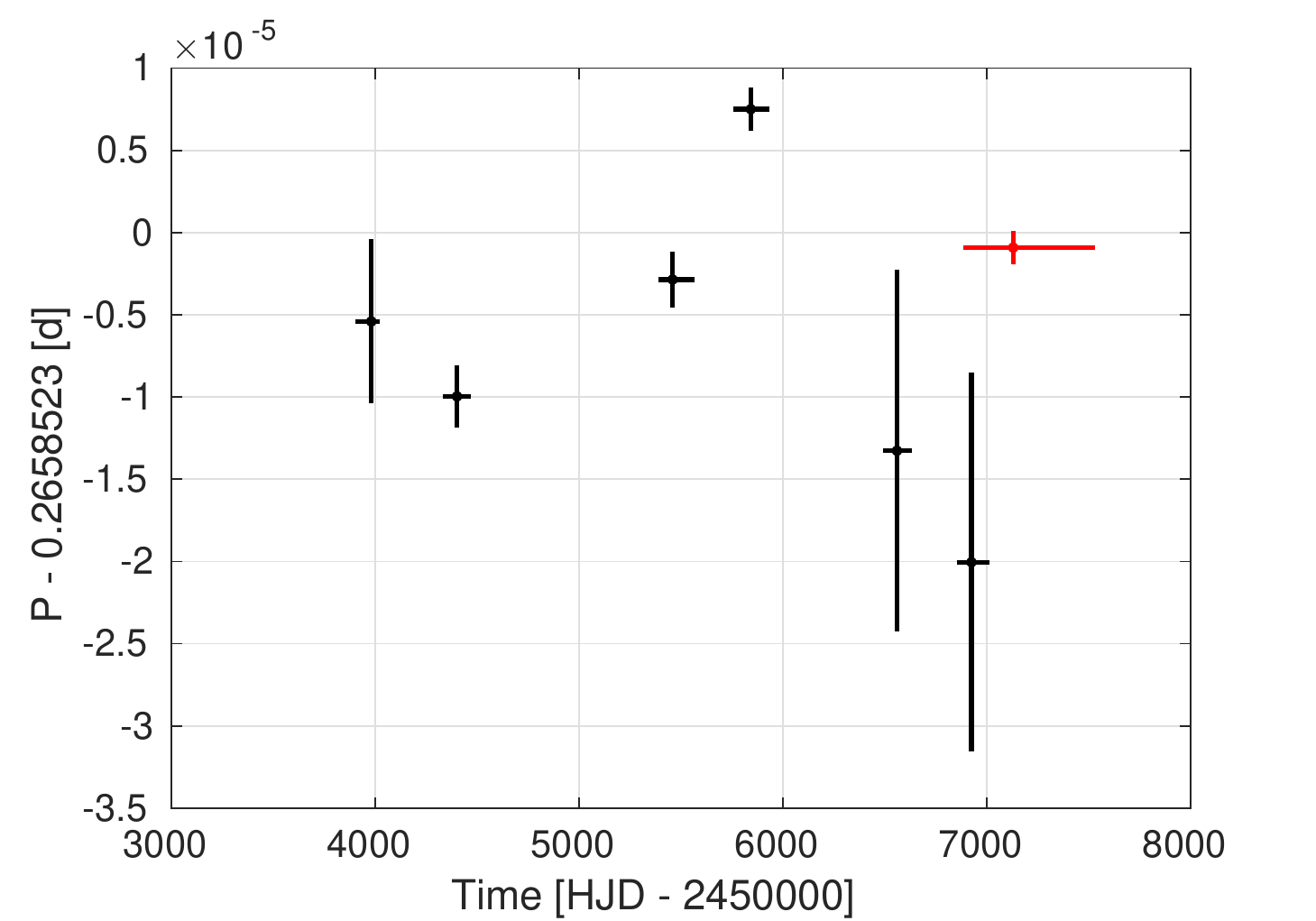}
	\caption{Period variability of SB 796. 
Independent period determination of six time segments. The horizontal bars represent the time span of each segment. The period uncertainty is presented by a vertical bar, and its location the time average of all data points in that section.
The rightmost red point presents the obtained period of {\it Gaia} G-band photometry (see below). 
 The period scatter reveals a probable variability of about $ 3\times 10^{-6}$ d on time scale of $\sim$ 500 d.}
	\label{fig:Period_Stability}
\end{figure*}

We obtained six different periods with a scatter of $\sim 3\times 10^{-6}$ d on time scales of $\sim 500$ d. Using these periods, each weighted by its inverse variance, we calculated a weighted mean value of P =  0.2658523 d. Assuming a fixed period model for the obtained periods, we found $\chi^2 \sim 70$, which is 14 times bigger than the expected value, indicating significant period variability. This phenomena was previously analyzed by \citet{moskalik15}, pointing out that RRc variables often show period variations of their dominant radial mode. SB 796 period variations are plotted in Fig.~\ref{fig:Period_Stability}. 
For the purpose of completeness, the figure also presents an independently derived period for the {\it Gaia} G-band photometry, which is discussed below.

In order to estimate the uncertainty of the average period, we normalized the weights to get $\chi^2$ = 5 (which is the number of degrees of freedom), finding a mean photometric period of $P = 0.2658523 \pm 0.0000033$ d.

In the same way, we derived the harmonic coefficients $A_n$, the amplitude ratio $R_{n1} \equiv A_n/A_1$ and the phase difference $\phi_{n1} \equiv \phi_n - n\phi_1$ between the n-th and first harmonic for each band, using seasons 1--4 (5--6) for the $V_W$ (r') band. In order to derive the time of minimum flux $T_0$, we fitted every season with a 3-harmonics model (equation~(\ref{eq:harmonics})), this time, imposing the mean photometric period P = 0.2658523 d. The weighted mean values are listed in Table~\ref{table:Photometry}. 

\begin{table}
	\centering
	\begin{tabular}{l c c} 
		\hline
		Parameter 				& $V_W$ photometry & r' photometry \\ \hline
		P [day]        			& \multicolumn{2}{c}{0.2658523 $\pm$ 0.0000033}						\\  \hline
		$T_{0}$  [HJD] 			&  2453983.5984   $\pm$ 0.0027		&  2453983.5697   $\pm$ 0.0071  	\\
		$A_{1}$     [mag] 		&  0.1109 		  $\pm$ 0.0016  	&  0.0592  		  $\pm$ 0.0027		\\
		$A_{2}$     [mag] 		&  0.01144 		  $\pm$ 0.00043  	&  0.00737 		  $\pm$ 0.00090		\\
		$A_{3}$  	[mag] 		&  0.00223 		  $\pm$ 0.00043  	&  0.00240 		  $\pm$ 0.00065		\\
		$R_{21}$      			&  0.1032 		  $\pm$ 0.0025  	&  0.1250  		  $\pm$ 0.0094		\\
		$\phi_{21}$   			&  4.519 		  $\pm$ 0.042  		&  5.30 		  $\pm$ 0.30		\\
		$R_{31}$      			&  0.0202 		  $\pm$ 0.0036  	&  0.0409 		  $\pm$ 0.0092		\\
		$\phi_{31}$   			&  2.60 		  $\pm$ 0.21  		&  2.3 		      $\pm$ 1.8			\\
		RMS of residuals [mag] 	&  0.067							&  0.27							\\
		\hline
	\end{tabular}
	\caption{
		Parameters of the photometric analysis of WASP light curve. Note the large Root-Mean-Square (RMS) of residuals in the r' photometry relative to the $V_W$ photometry.
	}
	\label{table:Photometry}
\end{table}

We found a first-harmonic semi-amplitude of $\sim 0.11$ ($0.06$) mag in the WASP $V_W$  (r') broadband filter. The second (third) harmonic coefficient was 2 (2--3) orders of magnitude smaller for both bands, as seen in Table~\ref{table:Photometry}. For the amplitude ratio, $R_{n1} \equiv A_n/A_1$ and the phase difference between the n-th and first harmonic, $\phi_{n1} \equiv \phi_n - n\phi_1$, the table shows an insignificance difference between the two bands. 

\begin{figure*}
	\centering
	\includegraphics[width=16cm, height=14cm]{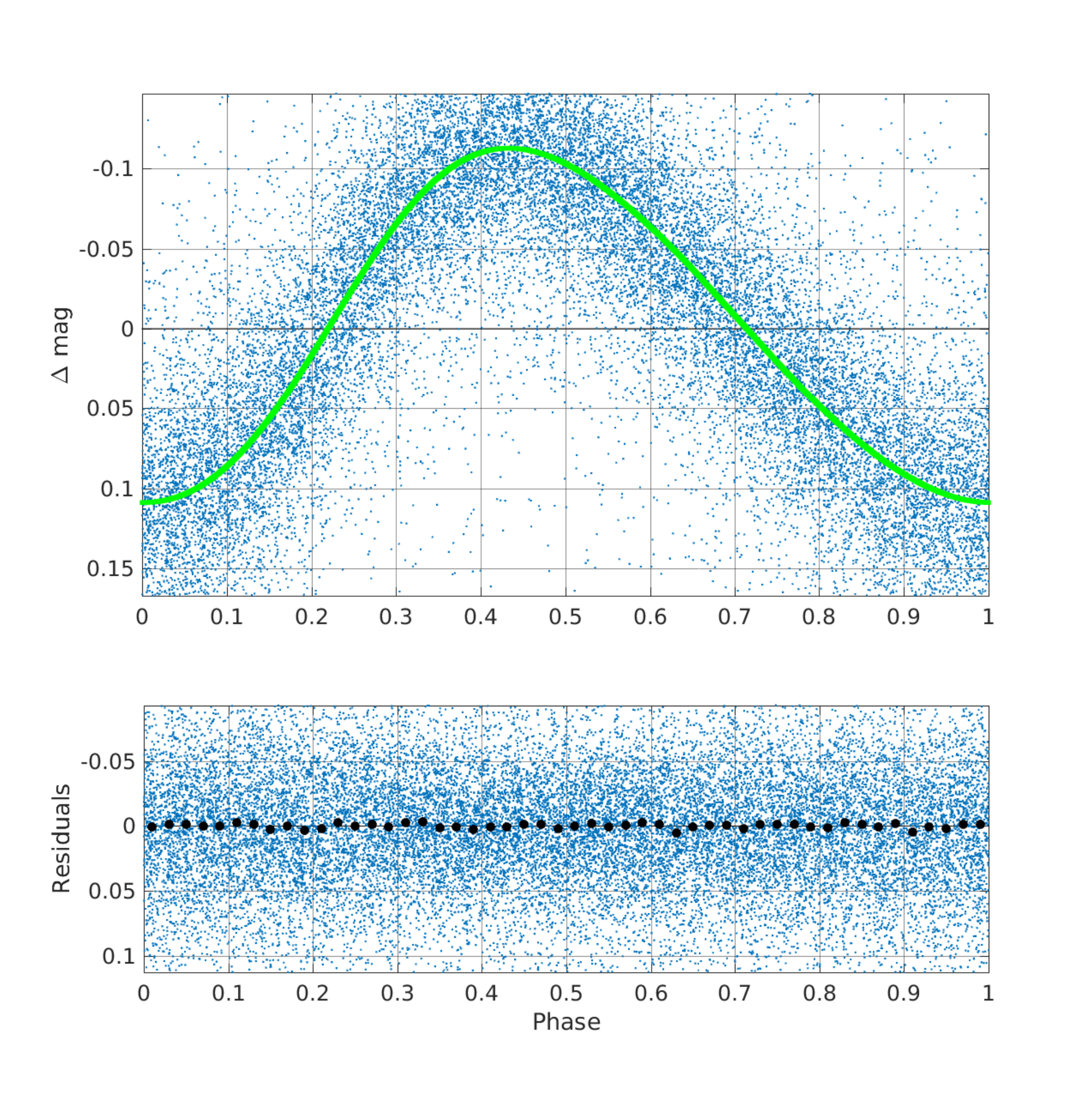}
	\caption{
		\textit{Top Panel}: Photometry of SB 796 in WASP $V_W$ band. Note the slight asymmetry of the modulation, with a peak at phase of $\sim~0.43$.  
		\textit{Bottom Panel}: Folded residuals. The black points show the residuals binned into 100 bins. The data (in both panels) are shown between the 5th-95th percentiles for clarity.}
	\label{fig:WASP_lc_folded_4S}
\end{figure*}

The light curve of SB 796 in WASP $V_W$ band is shown in Fig.~\ref{fig:WASP_lc_folded_4S}. The photometry was median subtracted and folded with a period of P = 0.2658523 d, with zero phase at $T_0 = 2453983.5985$ HJD, for which the 3-harmonics model gets its minimum. The model was calculated using the harmonics coefficients given in Table~\ref{table:Photometry}, plugged in equation~(\ref{eq:harmonics}).

\begin{figure*}
	\centering
	\includegraphics[width=15cm, height=10cm]{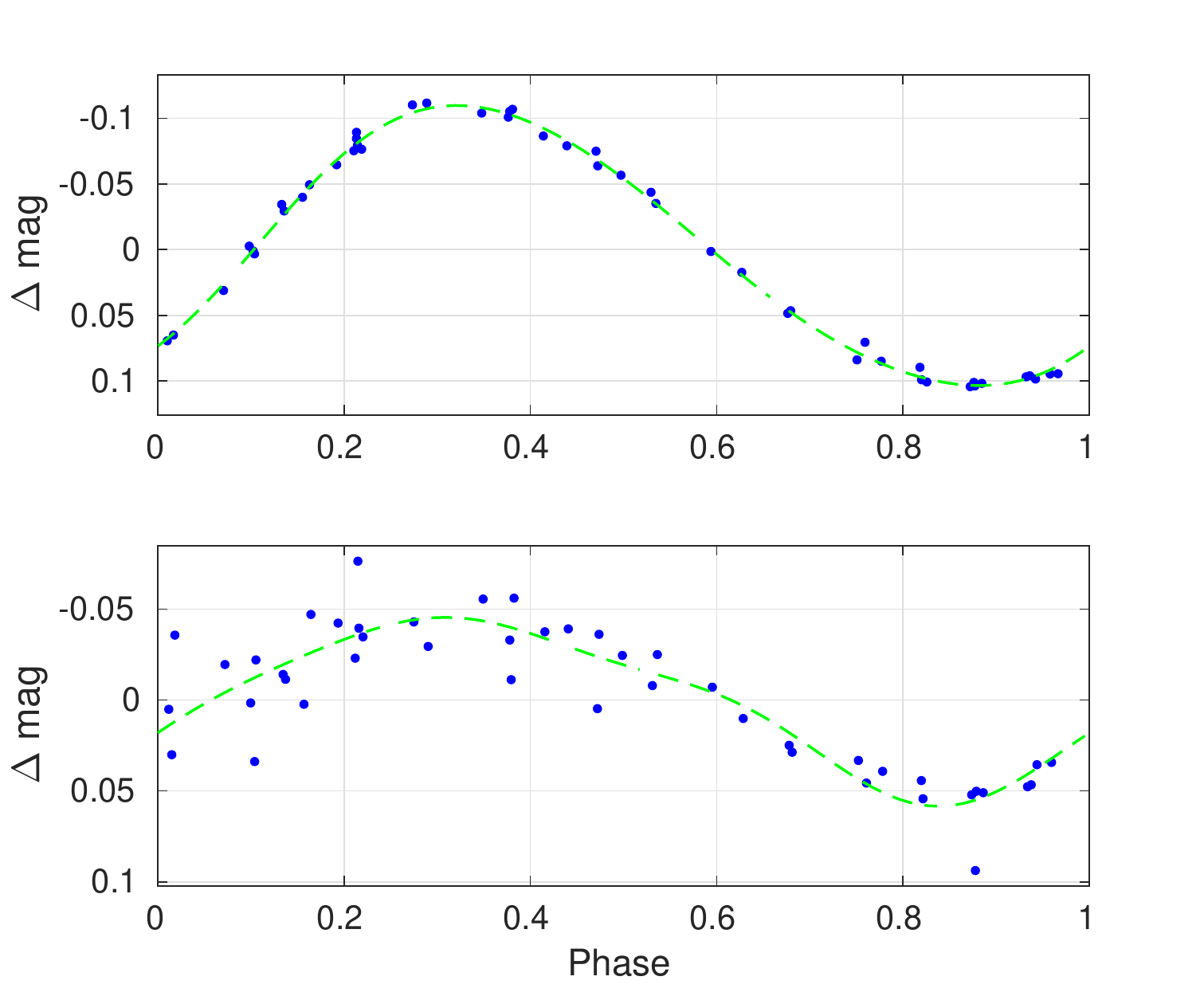}
	\caption{\textit{Upper Panel:} {\it Gaia} G band photometry of SB 796, folded with a period of P = 0.2658523 d, with zero phase at $T_0$ = 2453983.5984 HJD (minimum $V_W$ brightness) with a 3-harmonic  model. Note the slight asymmetry, with a peak at phase of $\sim~0.37$ and a minimum at phase of $\sim 0.94$, which has a similar shape to that of $V_W$ folded light curve (see Fig.~\ref{fig:WASP_lc_folded_4S}). \textit{Lower Panel:}  {\it Gaia} $G_{BP}-G_{RP}$ color of SB 796, which is the difference of integrated BP and RP spectro-photometric measurements, folded with the same period and phase, with a 3-harmonic model.
}
	\label{fig:GAIA_lc_folded}
\end{figure*}

{\it Gaia} DR2 \citep{clementini18} classifies SB 796 as an RRc variable ({\it Gaia} parameters are given in Table~\ref{table:GAIA}) and therefore includes 49, 46 and 46 measurements of the G, $G_{BP}$ and $G_{RP}$ bands, respectively, obtained in 2014--2016\footnote{https://gea.esac.esa.int/archive}. We used the photometric period found by the WASP data to fit the G band and the $G_{BP}-G_{RP}$ color with a 3-harmonic model, as illustrated in Fig.~\ref{fig:GAIA_lc_folded}. 

%

The color modulation $G_{BP}-G_{RP}$ correlates with the G band modulation, with the bluest color occurring at maximal flux, supporting the idea that the RR Lyrae modulation is associated with temperature modulation rather than radius modulation.

\newpage

\section{Spectroscopic observations}
\label{sec:RV}

To follow the stellar RV modulation, SB 796 was first observed 16 times at SAAO during 2017 December 6--19 with the SpUpNIC spectrograph and the G4 grating on the 1.9-m telescope \citep{crause16} with an exposure time of 450 s. Later it was observed  additional 9 times during 2017 December -- 2018 January  with the CORALIE spectrograph at the 1.2-m Euler Swiss telescope \citep{queloz00} with an exposure time of 1 h.

The spectra from the two observatories were analyzed with our \textit{UNICOR} software \citep{engel17}. First, we used \textit{UNICOR} to search the PHOENIX \citep{husser13} library\footnote{http://phoenix.astro.physik.uni-goettingen.de} for a template  that best fitted the spectra, using six CORALIE orders (16, 19, 24, 34, 40 and 65), which were of high signal-to-noise ratio (S/N), free of telluric lines and contained sufficient spectral information. We found the best fitted template of each order. Averaging the $T_{eff}$, logg and $[Fe/H]$ of the six templates yielded the parameters, with their uncertainties, of the adopted template, given in Table~\ref{table:Template}. 

\begin{table}
	\centering
	\begin{tabular}{l c} 
		\hline
		Parameter 		& Value \\ \hline
		Teff [K]     	& 8200  $\pm$  600  \\ 
		logg      		& 3.62  $\pm$  0.25 \\ 
		$[Fe/H]$ [dex]	& -1.6  $\pm$  0.5  \\
		\hline
	\end{tabular}
	\caption{Parameters of the adopted template with their uncertainties.}
	\label{table:Template}
\end{table}

\begin{figure}
	\centering
	\includegraphics[width=18cm, height=18cm]{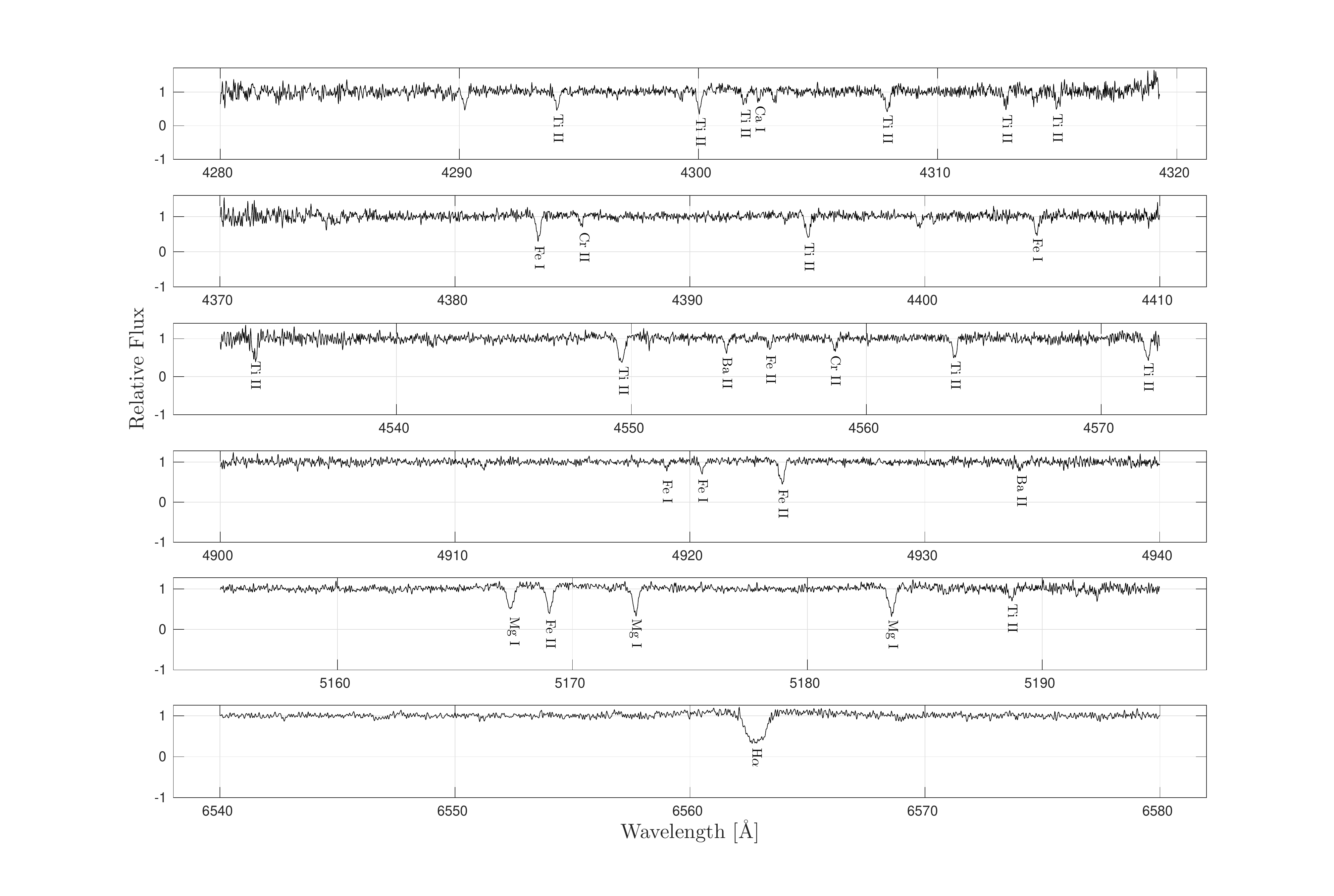}
	\caption{ De-blazed, rest-frame shifting and co-added spectra of SB 796 for the six CORALIE orders. }
	\label{fig:CORALIE_spec}
\end{figure}


The selected orders are shown in Fig.~\ref{fig:CORALIE_spec}. Each order was filtered with a low-pass window to approximate the instrumental response function of the spectrograph (the blaze). We divided the observed spectrum with the blaze to obtain the de-blazed spectrum. The figure presents a mean spectrum for each order, obtained by co-adding the de-blazed spectra, after shifted to the rest-frame of the star, using the derived velocity. The spectra are rich in lines of iron-group ions and other heavy metals, typical of RRc variables, as was previously shown by  \citet{govea14}.

We cross correlated each observed spectrum with the adopted template to obtain the corresponding RV, using either the SpUpNIC single order, or the six selected orders of CORALIE to yield one RV per exposure by \textit{UNICOR}. One CORALIE observation yielded low S/N spectrum and thus was excluded from the analysis. Table~\ref{table:RV_data} lists the timing of the observations, the RV measurements and their estimated uncertainties. The SpUpNIC RVs were assigned a 10 km/s uncertainty each, a typical scatter obtained in radial-velocity measurements of RV standards, performed with SpUpNIC during the current campaign.

\begin{table}
	\centering
	\begin{tabular}{l l l} 
		\hline
		Time [JD - 2450000]   & RV [km/s]   & $\sigma$ [km/s] \\ 
		\hline
		8094.3764 & 253 & 10 \\
		8094.4782 & 250 & 10 \\
		8094.4856 & 261 & 10 \\
		8095.3318 & 258 & 10 \\
		8096.3039 & 275 & 10 \\
		8097.34   & 266 & 10 \\
		8097.4162 & 272 & 10 \\
		8098.2932 & 264 & 10 \\
		8098.4742 & 259 & 10 \\
		8099.2799 & 264 & 10 \\
		8099.4663 & 250 & 10 \\
		8100.2817 & 264 & 10 \\
		8101.36   & 263 & 10 \\
		8102.3059 & 246 & 10 \\
		8106.3003 & 253 & 10 \\
		8107.3032 & 250 & 10 \\
		\hline
		8112.5502 & 253.63 & 0.61 \\
		8123.5603 & 243.12 & 0.87 \\
		8123.6448 & 250.93 & 0.68 \\
		8125.5706 & 253.76 & 0.98 \\
		8126.6367 & 254.9  & 1.6  \\
		8127.5641 & 243.3  & 1.1  \\
		8127.6282 & 249.66 & 0.94 \\
		8162.5373 & 254.2  & 2.3  \\
		\hline
	\end{tabular}
	\caption{
Derived radial-velocities of SpUpNIC (upper part) and CORALIE (lower part).
}
	\label{table:RV_data}
\end{table}

\begin{figure*}
	\centering
	\includegraphics[width=12cm, height=9cm]{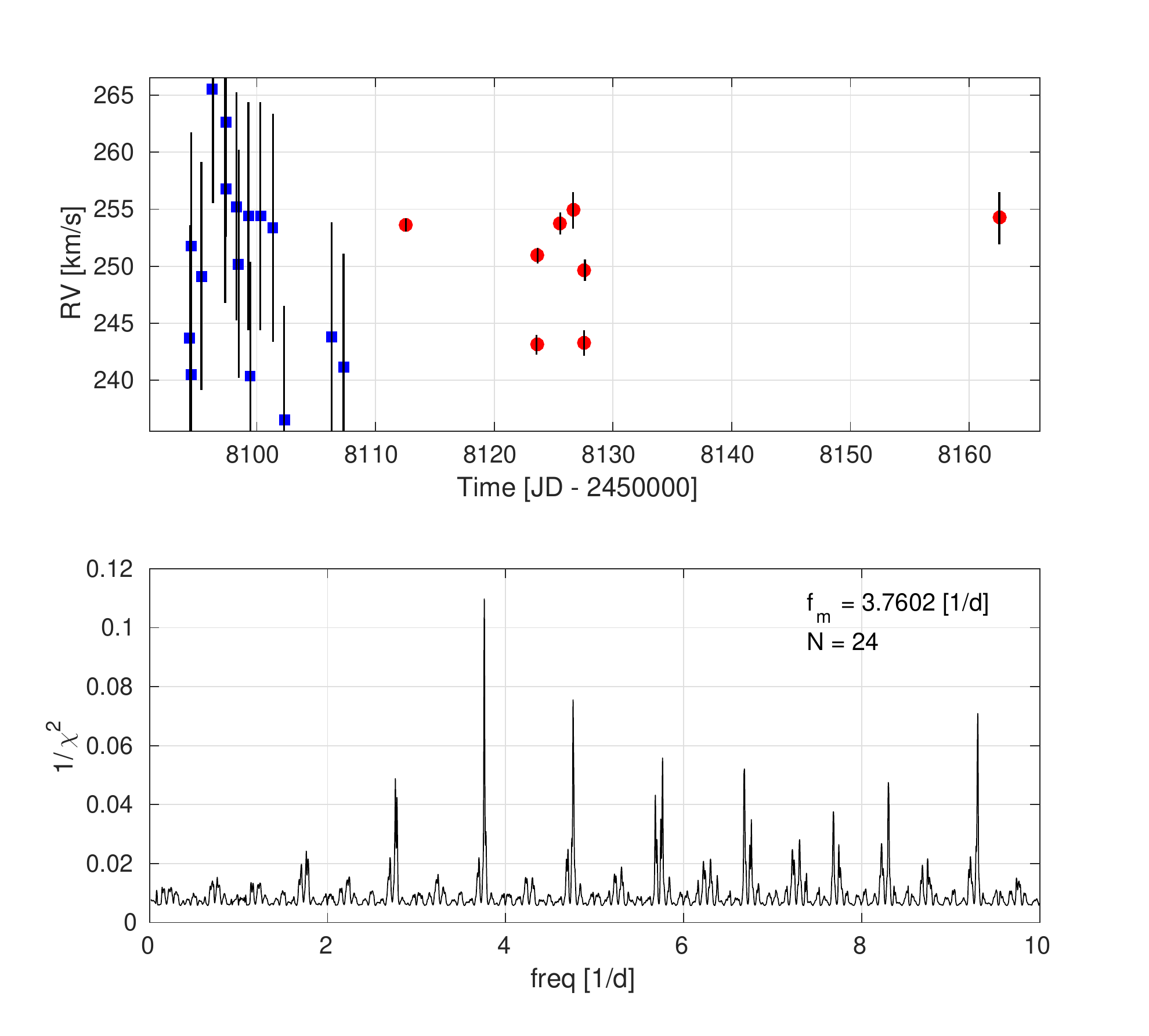}
	\caption{\textit{Top Panel}: RV as a function of time. Blue squares (red points) represent SpUpNIC (CORALIE) data. The SpUpNIC RVs were shifted by a $\gamma$ difference of $-9.65$ km/s, to put them on the CORALIE velocity system. \textit{Bottom Panel}: $\chi^2$ periodogram of a 1-harmonic fitting of the combined RV data as a function of frequency. A clear peak was found at a period of $0.265951 \pm 0.000072$ d. The other peaks are aliasing of the main peak due to the window function of the measurements.}
	\label{fig:SB796_RV_PS}
\end{figure*}

We derived a $\chi^2$ periodogram by performing a single harmonic fitting of the form 

\begin{equation}
 \gamma_{\scalebox{0.4}{\rm 1,2}} 
+ K_{\scalebox{0.4}{RV}}
\cos\big[
\frac{2\pi}{P_{\scalebox{0.4}{\rm RV}}}
(t-T_{0,\scalebox{0.4}{RV}})\big]
\end{equation}
to the combined RV sets of the two telescopes, while allowing for a constant shift between the two ($\gamma_i$ is the averaged line-of-sight velocity for each set). The analysis yielded a clear peak, as shown in Fig.~\ref{fig:SB796_RV_PS}, at $\sim$ 0.266 d. Using normalized uncertainties (so that each set has a reduced $\chi^2_{\rm red}=1$) we found a period of $ 0.265951 \pm 0.000072$ d, consistent with the photometric period.

Finally, we performed a single harmonic fitting to the CORALIE dataset, this time imposing the more accurate photometric period. The RV uncertainties were normalized to get a reduced $\chi^2_{\rm red}=1$ and can be found by dividing the CORALIE errors, given in Table~\ref{table:RV_sol}, by 2.7.

\begin{figure*}
	\centering
	\includegraphics[width=12cm, height=12cm]{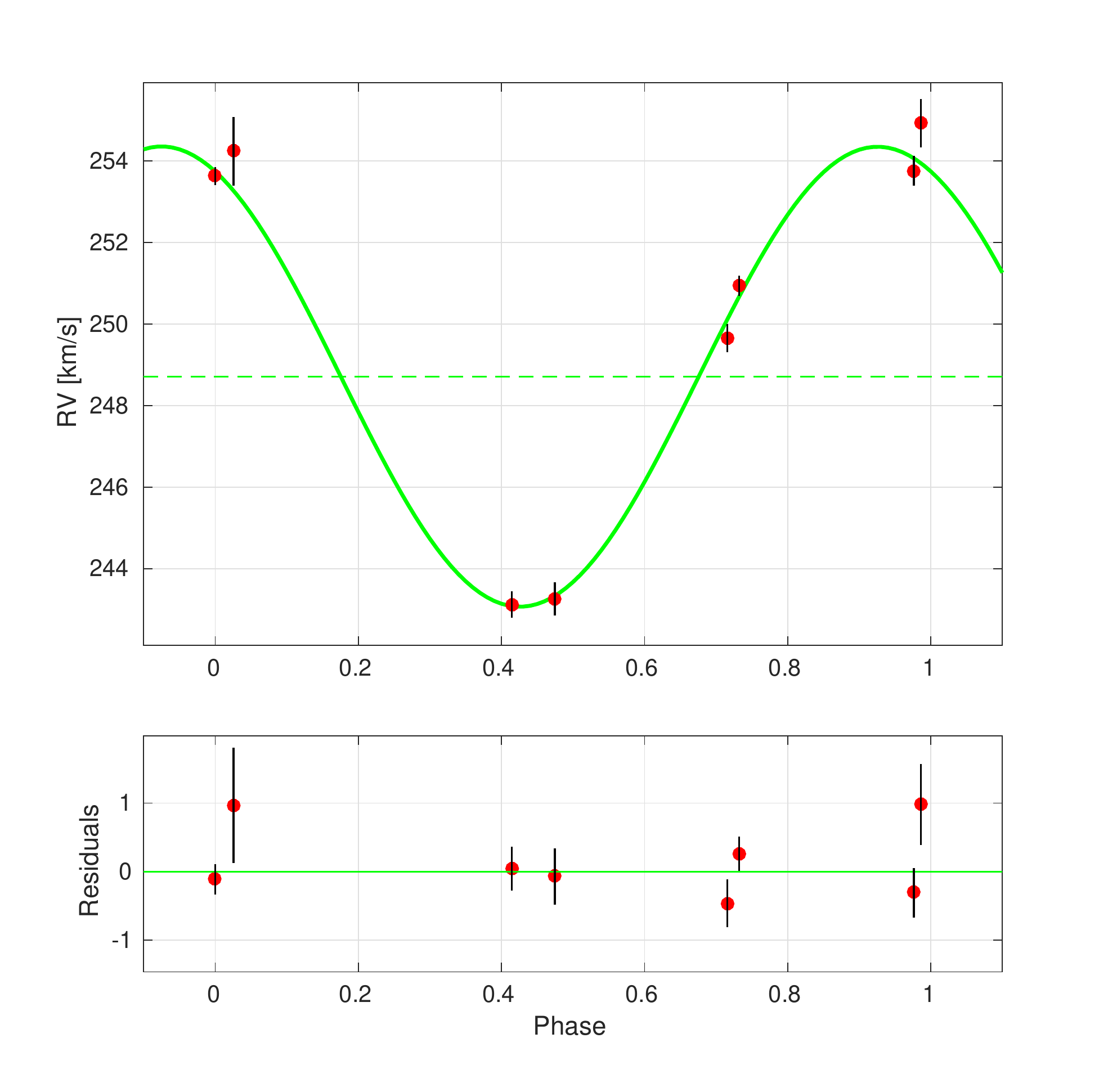}
	\caption{\textit{Top Panel}: Fitted RV modulation, folded with the photometric period of P = 0.2658523 d, with adopted zero phase at $T_0 = 2453983.5984$ HJD, for which the minimum $V_W$ brightness occurs. The green dashed line presents the average line-of-sight velocity. Minimal RV occurs at phase of $\sim$ 0.43. \textit{Bottom Panel}: Residuals as a function of time.}
	\label{fig:SB796_RV}
\end{figure*}

The fitted model is plotted in Fig.~\ref{fig:SB796_RV} and its parameters are listed in Table~\ref{table:RV_sol}. The phase of minimum RV coincided with the phase of maximum flux, up to $\sim 0.1$ cycle. Such coincidence was first noticed by \citet{sanford30} for two classical cepheids, and was verified also for RR Lyrae stars \citep{sneden17}.

\begin{table}
	\centering
	\begin{tabular}{l c} 
		\hline
		Parameter 						& Value \\ \hline
		$K_{\scalebox{0.5}{RV}}$ [km/s] & 5.64 $\pm$ 0.12  			   \\ 
		$\gamma$ [km/s]     			& 248.707 $\pm$ 0.088 		   \\
		$T_{0}$  [JD] 					& 2458100.03535 $\pm$ 0.00092  \\
		\hline
	\end{tabular}
	\caption{Parameters of the RV solution of SB 796 with the more accurate photometric period P = 0.2658523 d, where $T_{0}$ is the time of maximum RV.}
	\label{table:RV_sol}
\end{table}

\newpage

\section{Stellar classification and physical properties}
\label{sec:classification}

The combination of photometric and RV modulations indicates that SB 796 is a variable star with a period of $\sim 0.266$ d. 
The spectral analysis presented above 
suggested an effective temperature of $T_{\rm eff}= 8200 \pm 600$K and quite a low metallicity, typical of an RR Lyrae star.

\begin{figure*} 
	\centering
	\includegraphics[width=16cm, height=6cm]{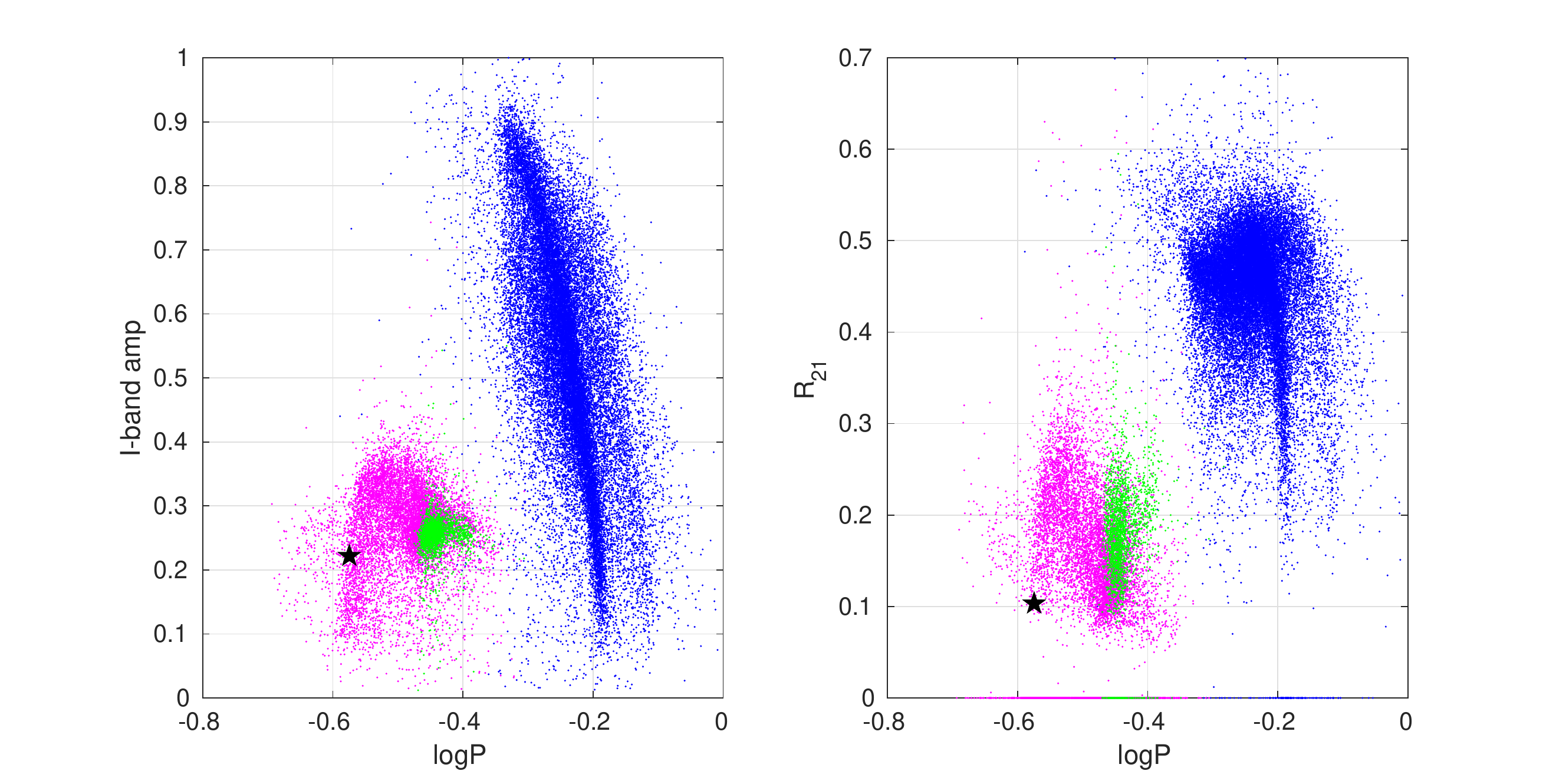}
	\caption{
		\textit{Left Panel}: Period amplitude diagram for RR Lyr in the LMC \citep{soszy09, soszy16}. Blue, magenta and green points show RRab, RRc and RRd stars, respectively. The $V_W$ light-curve parameters of SB 796 are presented by a star-shape point. \textit{Right Panel}: Fourier parameter $R_{21}$ as a function of logP for the different RR Lyrae classes in the LMC \citep{soszy09, soszy16}. $R_{21} \equiv A_2/A_1$ is the amplitude ratio of second to first harmonics \citep{simon81}.
}
	\label{fig:SB796_R21-PAD}
\end{figure*}

RR Lyrae stars can be sub-classified by the shape of their light curves, as done by \citet{soszy03,soszy09,soszy16} for the OGLE \citep{udalski15} LMC variables. They showed that the different RR Lyrae populations are well separated in diagrams of the light curve Fourier-coefficients ratio versus periods and in the period-amplitude diagram, sometimes called the Bailey diagram \citep{bailey1902}. The location of our system in these diagrams classifies SB 796 as an RRc variable, as clearly seen in 
Fig.~\ref{fig:SB796_R21-PAD}.

Based on its light-curve shape and position on the {\it Gaia} CMD  \citep{eyer18}, \citet{clementini18} classified SB 796 as an RRc Lyrae star, corroborating our result. 
The {\it Gaia} data of SB 796 is summarized in Table~\ref{table:GAIA}.

\begin{table}
	\centering
	\begin{tabular}{l c} 
		\hline
		Parameter 				  & Value \\ \hline
		source id            	  & 4958223672295639040           \\
		phot variable flag        & VARIABLE            		  \\
		best classification       & RRc                			  \\
		Period [d]                & 0.265853196 $\pm$ 0.000000059 \\
		\hline 
		band					  & G 							  \\
		num clean epochs          & 48                			  \\
		average signal [mag]      & 13.386382 $\pm$ 0.000065      \\
		peak to peak modulation [mag]      & 0.21461 $\pm$ 0.00030         \\
		$R_{21}$                  & 0.0915 $\pm$ 0.0015        	  \\
		$\phi_{21}$               & 4.369 $\pm$ 0.012        	  \\
		\hline
		band					  & $G_{BP}$ 					  \\
		num clean epochs          & 46                			  \\
		average signal [mag]      & 13.49126 $\pm$ 0.00028        \\
		peak to peak modulation [mag]      & 0.2512 $\pm$ 0.0015           \\
		\hline 
		band					  & $G_{RP}$ 					  \\
		num clean epochs          & 45                			  \\
		average signal [mag]      & 13.15588 $\pm$ 0.00028       \\
		peak to peak modulation [mag]      & 0.1558 $\pm$ 0.0018           \\
		\hline
		$G_{BP}-G_{RP}$ [mag]     & 0.33562 $\pm$ 0.00040                		  \\
		$G_{BP}-G$ [mag]          & 0.10286 $\pm$ 0.00029                		  \\
		$G-G_{RP}$ [mag]          & 0.23275 $\pm$ 0.00029                		  \\
		Teff [K]            	  & 8521.50            			  \\
		Teff percentile lower [K] & 8071.0            			  \\
		Teff percentile upper [K] & 9004.5            			  \\
		\hline
	\end{tabular}
	\caption{
{\it Gaia} parameters of SB 796. The color uncertainties were calculated using the average signal uncertainty of each band.
} 
	\label{table:GAIA}
\end{table}

\begin{figure*}
	\centering
	\includegraphics[width=12cm, height=10cm]{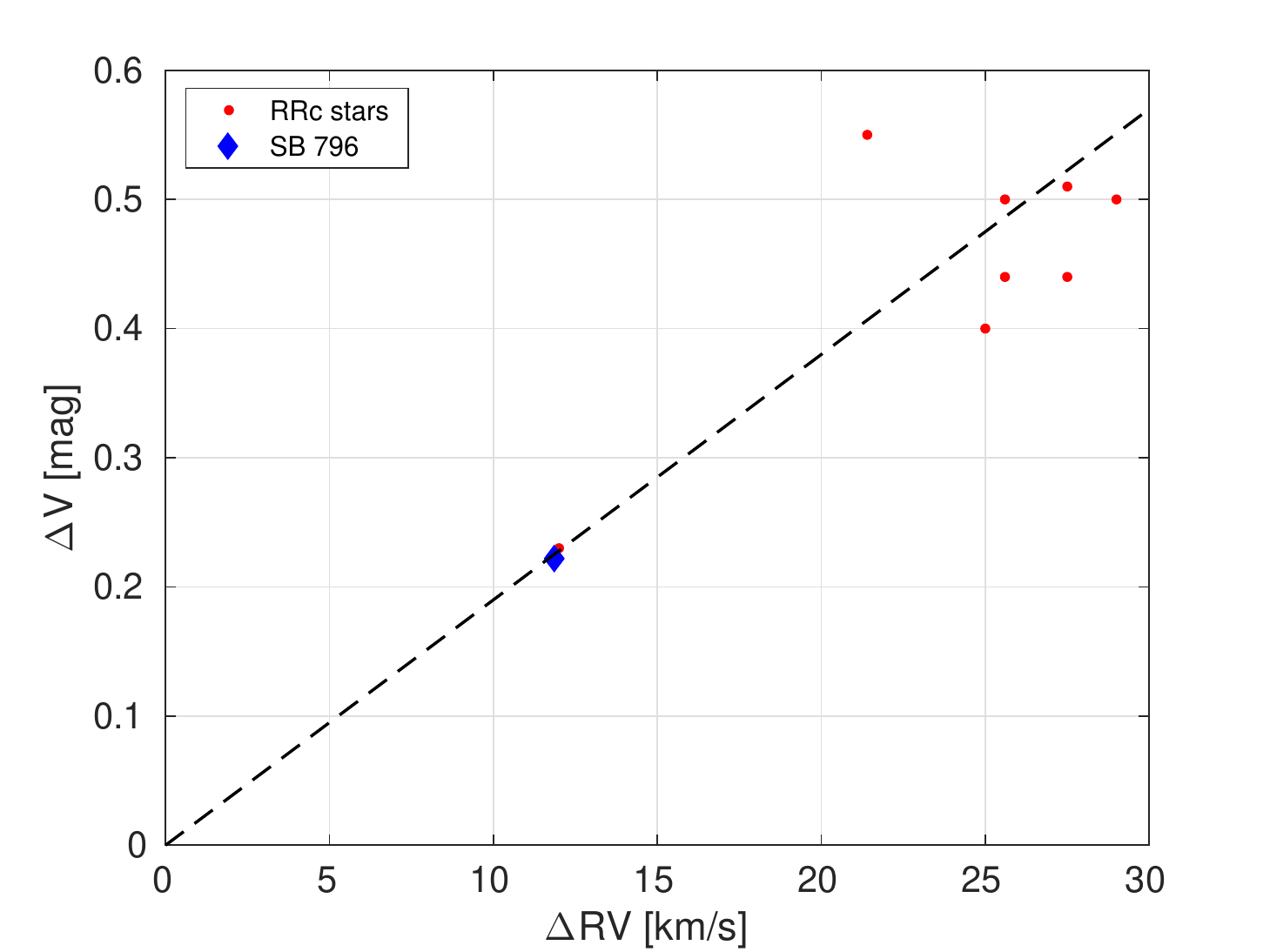}
	\caption{
Visual-light as a function of RV amplitudes. The red points present RRc variables studied by Sneden and fitted with the dashed regression line. SB 796 is presented by a diamond-shape symbol; its uncertainties cannot be resolved due to their small values: $\Delta V_W = 0.0032$ [mag] and $\Delta RV = 0.37$ [km/s].
}
	\label{fig:SB796_Sneden}
\end{figure*}

\citet{sneden17} studied 12 RRc variables that were chosen from the brightest RRc stars detected by the All Sky Automated Survey\footnote{available at http://www.astrouw.edu.pl/asas/?page=main} (ASAS) and derived a linear relation between their 
visual-light and RV amplitudes. With $V_W$ amplitude of 0.2227 $\pm$ 0.0032 [mag] and RV amplitude of 11.28 $\pm$ 0.37 [km/s], SB 796 almost coincides with the lowest point of \citet{sneden17},  as illustrated in Fig.~\ref{fig:SB796_Sneden}.

\newpage

%
\section{Galactic dynamics of SB 796}
\label{sec:galactic}

In order to study the Galactic dynamics of SB 796 we derived its current Galactic position, using its equatorial coordinates, parallax, proper motions and mean RV, listed in Table~\ref{table:SB796-loc}. {\it GAIA} parallax uncertainty of SB 796 is 0.02013 mas. However, as pointed out by \citet{lindegren18}, {\it GAIA} parallaxes are on the whole too small by a median value of $\sim$ 0.03 mas. Thus we adopted a parallax uncertainty of 0.02 mas.

We then used the \texttt{galpy} package\footnote{http://github.com/jobovy/galpy} for Galactic motion calculations \citep{bovy15} to integrate the stellar orbit  for the past and into the future for 1 Gyr, adopting the \texttt{MWPotential2014} approximation for the Milky-Way gravitational field \citep{bovy13}. The adopted model of the Galaxy is given in Table~\ref{table:Gal-Model}.

\begin{table}
	\begin{center}
		\begin{tabular}{ l c } 
			\hline
			Parameter							& Value \\ \hline
			RA$^a$ [deg] 					    & 30.6212071184 $\pm$ 3.7 $\cdot$ $10^{-9}$    \\ 
			DEC$^a$ [deg]     					& -40.3739311288 $\pm$ 3.9 $\cdot$ $10^{-9}$   \\ 
			parallax$^a$ [mas]     				& 0.26 $\pm$ 0.02  \\ 
			pm in RA$^a$ [mas/yr]   			& 17.113 $\pm$ 0.024   \\
			pm in DEC$^a$ [mas/yr]   				& -0.499 $\pm$ 0.027   \\
			RV$^b$ [km/s]						& 248.44 $\pm$ 0.22    \\
			Gal position (X,Y,Z) [kpc]			& (8.242,-1.31,-3.70)  $\pm$ (0.019,0.11,0.30) \\
			Gal velocity (Vx,Vy,Vz) [km/s] 		& (229,-59,-138.1) $\pm$ (18,17,7.1) \\
			\hline \\
			$^a${\it Gaia} DR2 values \\
			$^b$CORALIE values \\
		\end{tabular} \\
		\caption{Equatorial coordinates, parallax, proper motions, mean radial-velocity and corresponding Galactic position and rest-frame velocity of SB 796.}
		\label{table:SB796-loc}
	\end{center}
\end{table}

\begin{table}
	\begin{center}
		\begin{tabular}{ l c } 
			\hline
			Parameter							& Value \\ \hline
			Buldge component$^a$ \\
			mass $ [10^{10}M_\odot$] 			& 0.5 $\pm$ 0.25	\\
			cut-off radius [kpc]				& 1.9				\\
			power-law exponent					& 1.8				\\
			\hline
			Stellar disk component$^a$ \\
			mass $ [10^{10}M_\odot$]			& 6.8 $\pm$ 0.5		\\
			horizontal scale-length [kpc]		& 3.0 $\pm$ 0.2		\\
			vertical scale-length [kpc]			& 0.28 $\pm$ 0.05	\\
			\hline
			Dark halo component$^a$ \\
			local dark matter density $[M_\odot/pc^3]$	& 0.008 $\pm$ 0.0025 \\
			radial scale-length [kpc]			& 16				\\
			\hline \\
			$^a$\citet{bovy15}, adopting relative uncertainties as in \citet{bovy13}. \\
		\end{tabular} \\
		\caption{Model components adopted for the Milky-Way.}
		\label{table:Gal-Model}
	\end{center}
\end{table}

\begin{figure*}
	\centering
	\includegraphics[width=16cm, height=12cm]{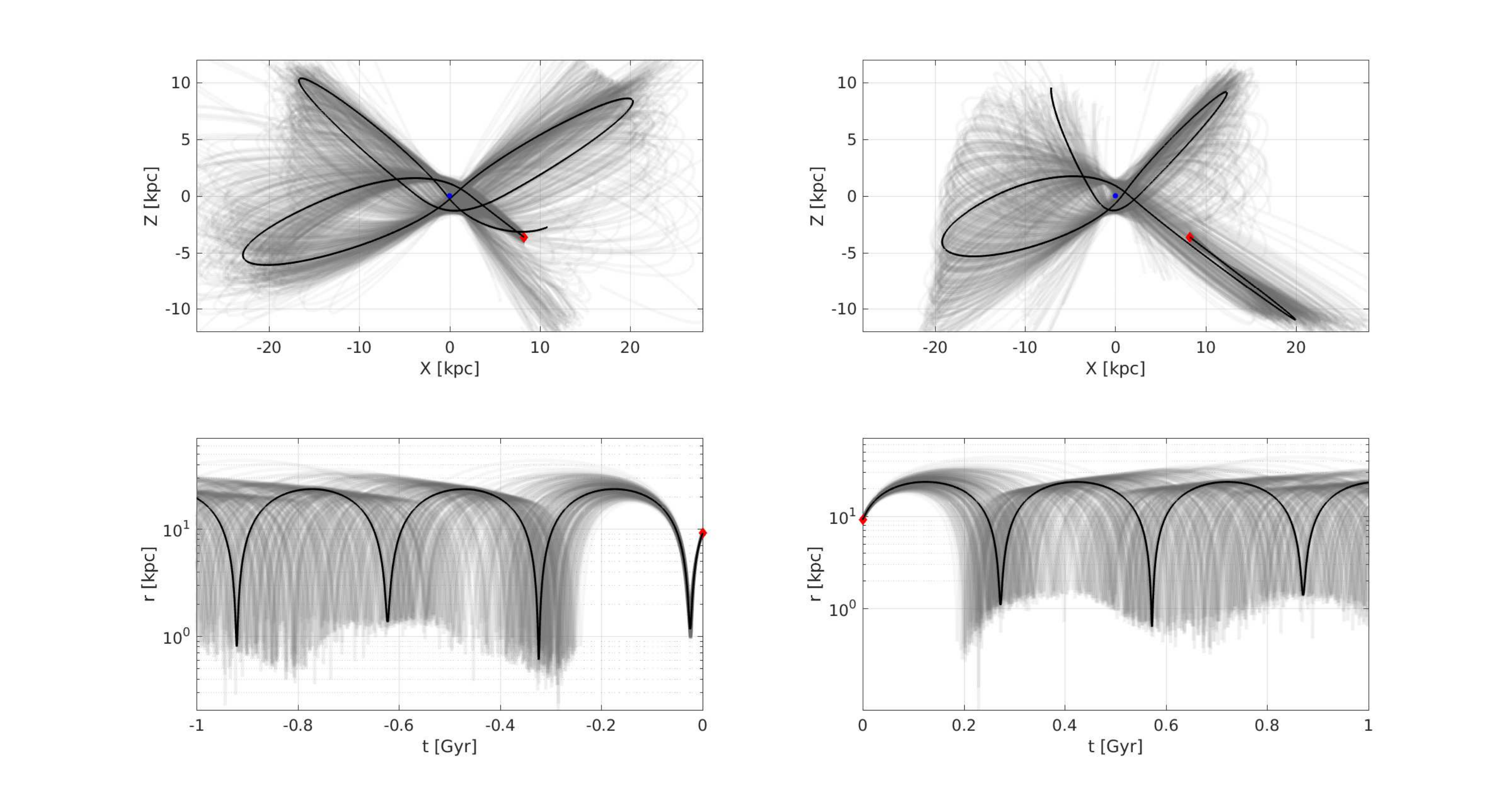}
	\caption{
SB 796 Galactic orbit in a left-handed Galactocentric rest frame. Each gray line corresponds to a possible track obtained by the simulation. The orbital motion is plotted in the upcoming 1 Gyr (right panels) and in the past 1 Gyr (left panels). The orbits in the x-z plane are presented in the upper panels, while lower panels show the galacto-centric distances as a function of time. The red diamond shows the present time position in all panels and the thick line presents the track obtained using the adopted Galactic parameters. 
}
\label{fig:SB796_orbit}
\end{figure*}

The resulting Galactic orbit is plotted in Fig.~\ref{fig:SB796_orbit}. The star moves on a Galactic radial orbit  that takes it up and down the plane to a scale height of $\sim 10$ kpc. The orbit crosses the neighborhood of the Galactic center once in $\sim 0.3$ Gyrs.  

The figure also shows an assortment of orbits with different Galactic initial conditions,
chosen at random from Gaussian distributions centered on the measured values of SB 796 with widths equal to the derived uncertainties, combined with uncertainties of the Milky-Way mass model parameters. We found that the actual SB 796 Galactic orbit could significantly deviate from the adopted path after (before) $\sim 0.1$ Gyrs. Calculating the orbital tracks for past and future times of 1 Gyr, with different initial conditions and mass profiles, yielded a minimal Galacto-centric distance of $r_{\rm min}= 0.6 \pm 0.2$ kpc. 

\newpage

%
\section{Summary}
\label{sec:summary}
%

Based on WASP photometric data, we have shown that SB 796 is an RRc variable with a period of  $P = 0.2658523 \pm 0.0000033$ d, semi-amplitude of $\sim$ 0.11 mag at optical wavelengths and a typical shape of the photometric modulation. We found a small but significant period variability of $3\times 10^{-6}$ d on time scales of $\sim 500$ d.

{\it Gaia} DR2 classification \citep{clementini18}, released when we were preparing our results for publication, corroborated our result. 
The {\it Gaia} color modulation, $G_{BP}-G_{RP}$, showed that maximum brightness occurs when the star is most blue, indicating that the modulation is associated with temperature modulation rather than radius modulation.

The pulsation is accompanied by a radial-velocity modulation with a semi-amplitude of about 6 km/s. The minimum of the RV, when the stellar surface facing us is moving towards us at maximum absolute velocity, occurs at the phase of maximum brightness, as was seen  by \citet[e.g.,][]{sneden17} in other RRc variables.
The RV and visual modulation amplitudes of SB 796 are also consistent with the derived linear relation found by \citet{sneden17}.
Detailed and accurate RV follow-up of the modulation can be used to confront and elaborate the theory of the RR Lyrae pulsation. 

SB 796 is presently located at $3.7 \pm 0.3 $ kpc below the Galactic plane, and its Galactic radial motion takes it up and down the plane to a scale height of $\sim 10$ kpc. During its motion, the star passes near the Galactic center, within $\sim 0.5$ kpc, about three times within 1 Gyr. During its $\sim10$ Gyrs life time, SB 796 therefore passed $\sim 30$ times near the Galactic center. Each passage substantially changes the characteristics of the stellar motion, as seen in Fig.~\ref{fig:SB796_orbit}. 

In the future, when {\it Gaia} DR3 and DR4 are available, we will be able to perform similar detailed studies of the many RRc variables discovered by the mission, and find out if they all share similar Galactic orbits. This could be a probe into their early dynamical history, when the  Galaxy was still young.

\section*{Acknowledgments}
%

We thank the referee Floor Van Leeuwen whose comments and suggestions are highly appreciated.
Our study uses observations made at the South African Astronomical Observatory, which is supported by the National Research Foundation of South Africa.
It also makes use of data from the first public release of the WASP data (Butters et al. 2010) as provided by the WASP consortium and services at the NASA Exoplanet Archive, which is operated by the California Institute of Technology, under contract with the National Aeronautics and Space Administration under the Exoplanet Exploration Program.
Finally, we have made use of data from the European Space Agency (ESA) mission {\it Gaia} (\url{https://www.cosmos.esa.int/gaia}), processed by the {\it Gaia} Data Processing and Analysis Consortium (DPAC, \url{https://www.cosmos.esa.int/web/gaia/dpac/consortium}). Funding for the DPAC has been provided by national institutions, in particular the institutions participating in the {\it Gaia} Multilateral Agreement.
This research was supported by Grant No.~2016069 of the United States-Israel Binational Science Foundation (BSF) and by the Israeli Centers for Research Excellence (I-CORE, grant No.~1829/12). 

\newpage

\bibliographystyle{mnras}
\bibliography{SB796_bib}
\end{document}